%% file: HIN-18-009_temp.tex
\pdfoutput=1

\documentclass[11pt,twoside,a4paper,cmspaper,final,collab]{cms-tdr}
\def\svnVersion{aa43977-D}\def\svnDate{2020/02/09}\def\cmsCernNoTag{CERN-EP-2019-102}\def\cmsCernDate{\today}\def\cmsMessage{"Published in Physics Letters B as \href{http://dx.doi.org/10.1016/j.physletb.2020.135328}{\doi{10.1016/j.physletb.2020.135328}.}"}

\begin{document}\cmsNoteHeader{HIN-18-009}

\hyphenation{had-ron-i-za-tion}
\hyphenation{cal-or-i-me-ter}
\hyphenation{de-vices}
\RCS$HeadURL$
\RCS$Id$

\newlength\cmsFigWidth
\ifthenelse{\boolean{cms@external}}{\setlength\cmsFigWidth{0.49\textwidth}}{\setlength\cmsFigWidth{0.6\textwidth}}
\newlength\cmsFigWidthHalf
\ifthenelse{\boolean{cms@external}}{\setlength\cmsFigWidthHalf{0.45\textwidth}}{\setlength\cmsFigWidthHalf{0.5\textwidth}}
\ifthenelse{\boolean{cms@external}}{\providecommand{\cmsLeft}{upper\xspace}}{\providecommand{\cmsLeft}{left\xspace}}
\ifthenelse{\boolean{cms@external}}{\providecommand{\cmsRight}{lower\xspace}}{\providecommand{\cmsRight}{right\xspace}}
\providecommand{\PbPb}{\ensuremath{\text{PbPb}}\xspace}
\providecommand{\AuAu}{\ensuremath{\text{AuAu}}\xspace}
\newcommand{\LcDratio}{\ensuremath{\PcgLp/ \PDz}\xspace}
\newcommand{\Lcdecay}{\ensuremath{\PcgLp\to\Pp\PKm\Pgpp}\xspace}
\newcommand{\RAA}{\ensuremath{R_{\mathrm{AA}}}\xspace}
\newcommand{\taa}{\ensuremath{T_{\mathrm{AA}}}\xspace}
\newcommand{\absy}{\ensuremath{\abs{y}}\xspace}
\newcommand{\Ncoll}{\ensuremath{N_\text{coll}}\xspace}
\newcommand{\Npart}{\ensuremath{N_\text{part}}\xspace}
\newcommand{\abseta}{\ensuremath{ \abs{\eta}}\xspace}
\newcommand{\Raap}{\ensuremath{R_{\mathrm{AA}}^{\text{prompt}}}}
\newcommand{\Raanon}{\ensuremath{R_{\mathrm{AA}}^{\text{nonprompt}}}}
\newcommand{\Raaratio}{\ensuremath{\Raanon/ \Raap}\xspace}

\cmsNoteHeader{HIN-18-009}
\title{Production of $\PcgLp$ baryons in proton-proton and lead-lead collisions at $\sqrtsNN=5.02\TeV$}

\date{\today}

\abstract{
The transverse momentum ($\pt$) spectra of inclusively produced $\PcgLp$ baryons are measured via the exclusive decay channel $\Lcdecay$ using the CMS detector at the LHC. Spectra are measured as a function of transverse momentum in proton-proton ($\Pp\Pp$) and lead-lead ($\PbPb$) collisions at a nucleon-nucleon center-of-mass energy of 5.02\TeV.
The measurement is performed within the $\PcgLp$ rapidity interval $\absy<1$ in the $\pt$ range of 5--20\GeVc in $\Pp\Pp$ and 10--20\GeVc in $\PbPb$ collisions. The observed yields of $\PcgLp$ for $\pt$ of 10--20\GeVc suggest a suppression in central $\PbPb$ collisions compared to $\Pp\Pp$ collisions scaled by the number of nucleon-nucleon (NN) interactions. The $\LcDratio$ production ratio in $\Pp\Pp$ collisions is compared to theoretical models. In $\PbPb$ collisions, this ratio is consistent with the result from $\Pp\Pp$ collisions in their common $\pt$ range.
}

\hypersetup{
pdfauthor={CMS Collaboration},
pdftitle={Production of Lambda(c) baryons in proton-proton and lead-lead collisions at sqrt(sNN) = 5.02 TeV},
pdfsubject={CMS},
pdfkeywords={CMS, physics, Lambda(c) baryons, nuclear modification factor, heavy flavor}}

\maketitle

\section{Introduction}
Measurements of heavy-quark production provide unique inputs in understanding the parton energy loss and the degree of thermalization in the quark-gluon plasma (QGP)~\cite{QGP} formed in high energy heavy ion collisions.
Compared to light quarks, different energy loss mechanisms~\cite{energyloss} are expected to dominate the interaction between heavy quarks and the medium.
Besides the in-medium interactions, a detailed study of the hadronization process is critical for the interpretation of experimental data. In relativistic heavy ion collisions, in addition to the fragmentation process present in proton-proton ($\Pp\Pp$) collisions, hadron production can also occur via coalescence, where partons combine with each other while traversing the QGP medium or at the phase boundary~\cite{Greco_introduction,Greco_follow}.
At high transverse momentum ($\pt\gtrsim6\GeVc$), the probability of coalescence is reduced, and therefore the hadronization process is expected to be dominated by fragmentation.
In the intermediate $\pt$ region ($2\lesssim\pt\lesssim6\GeVc$), a significant enhancement of the baryon-to-meson ratio is observed in heavy ion collisions for hadrons with up, down, or strange quarks~\cite{light_hadron_1, light_hadron_3}. This enhancement, and its dependence on centrality (\ie, the degree of overlap of the two colliding nuclei) can be explained in a scenario with hadronization via coalescence. Furthermore, elliptic flow, the second Fourier component of the azimuthal distribution of emitted particles, is found to roughly scale with the number of constituent quarks in the $\pt$ range of 2--5\GeVc at RHIC~\cite{charged_v2v3}, an observation which is also consistent with the expectation for coalescence. 

A significant contribution of coalescence to the hadronization of charm quarks from the QGP medium is supported by various measurements of charmonium and open charm production at RHIC and LHC energies~\cite{PHENIX_JPSI,ALICE_JPSI_502,ALICE_JPSI_276,update_D0Raa,newD0result,CMS_D0v2v3,HIN16001,ALICE_D0v2v3,ALICE_D0_RAA}.
One such observable is the nuclear modification factor, $\RAA$, which is the ratio of the yield in heavy ion collisions to that in $\Pp\Pp$ collisions scaled by the number of nucleon-nucleon (NN) interactions.
At RHIC, the $\RAA$ for $\JPsi$ mesons with $\pt\leq7\GeVc$ produced in $\AuAu$ collisions decreases significantly from peripheral to central collisions~\cite{PHENIX_JPSI}. In contrast, in higher energy $\PbPb$ collisions at the LHC, the $\JPsi$ $\RAA$ has a much smaller centrality dependence~\cite{ALICE_JPSI_502,ALICE_JPSI_276}. The difference between the $\AuAu$ and $\PbPb$ results can be explained by a larger coalescence probability in $\PbPb$ collisions because of the larger number of produced charm and anti-charm quarks at the higher center-of-mass energy.
For $\PDz$ meson production in $\AuAu$ collisions, $\RAA$ is observed to increase with $\pt$ up to 1.5\GeVc and decrease with $\pt$ from 2 to 6\GeVc, an effect that can be qualitatively reproduced by models involving coalescence~\cite{update_D0Raa,newD0result}. At the LHC, the measurements of $\PDz$ $\RAA$ and $\PDz$ azimuthal anisotropy~\cite{CMS_D0v2v3,HIN16001,ALICE_D0v2v3,ALICE_D0_RAA} are well explained by models involving coalescence.
The relative coalescence contribution to baryon production is expected to be more significant than for mesons because of their larger number of constituent quarks.
In particular, models involving  coalescence of charm and light-flavor quarks  predict a large enhancement in the $\LcDratio$ production ratio in heavy ion collisions relative to $\Pp\Pp$ collisions and also predict that the enhancement has a strong $\pt$ dependence~\cite{Oh09, Lee, ghosh,Greco_publication}. Comparison of $\PcgLp$ baryon production in $\Pp\Pp$ and lead-lead ($\PbPb$) collisions can thus shed new light on understanding heavy-quark transport in the medium and heavy-quark hadronization via coalescence. All discussions of $\PcgLp$ and $\PDz$ also include the corresponding charge conjugate states.

Recently, the production of $\PcgLp$ baryons for a variety of collision configurations has been measured in a similar \pt range by the LHC experiments ALICE and LHCb in the central and forward rapidity regions, respectively~\cite{ALICE_pp,ALICE_PbPb,LHCb_pp,LHCb_result}.
Both experiments measured the $\PcgLp$ \pt-differential cross sections in $\Pp\Pp$ collisions at a center-of-mass energy of $\sqrt{s}=7\TeV$ and compared them to theoretical predictions using the next-to-leading order Generalized Mass Variable Flavor Number Scheme~\cite{GMVFNS}. The LHCb results for the rapidity range $2.0<y<4.5$ were found to be compatible with theory~\cite{LHCb_pp}, while the ALICE values for $\absy<0.5$ were larger than the predictions~\cite{ALICE_pp}. The ALICE experiment also reported $\LcDratio$ production ratios in 7\TeV $\Pp\Pp$ collisions, as well as in proton-lead ($\Pp$Pb) and $\PbPb$ collisions at an NN center-of-mass energy of $\sqrtsNN=5.02\TeV$. The ALICE ratios from $\Pp\Pp$ and $\Pp$Pb collisions~\cite{ALICE_pp} were found to be above the corresponding LHCb values~\cite{LHCb_result} (however in different rapidity ranges), with the latter agreeing with theoretical predictions.
The ALICE $\LcDratio$ production ratio for $6<\pt<12\GeVc$ in $\PbPb$ collisions was measured to be larger than in $\Pp\Pp$ and $\Pp$Pb collisions~\cite{ALICE_PbPb}, and this difference can be described using a model involving only coalescence in hadronization~\cite{Greco_publication}. The ALICE measurements of the $\RAA$ of $\PcgLp$ baryons in $\Pp$Pb and $\PbPb$ collisions were found to be compatible with unity and less than unity, respectively, but have limited power to constrain models owing to large uncertainties~\cite{ALICE_pp,ALICE_PbPb}.

In this letter, we report measurements of inclusive $\PcgLp$ baryon production in $\Pp\Pp$ and $\PbPb$ collisions at high $\pt$ where inclusive refers to both prompt (directly produced in charm quark hadronization or from strong decays of excited charmed hadron states) and nonprompt (from $\Pb$ hadron decays) production. The data were collected at $\sqrtsNN=5.02\TeV$ in 2015 using the CMS detector. The $\PcgLp$ baryons are reconstructed in the central region ($\absy<1$) via the hadronic decay channel $\Lcdecay$.  The \pt spectrum and $\LcDratio$ production ratio are measured in the $\pt$ ranges 5--20 and 10--20\GeVc in $\Pp\Pp$ and $\PbPb$ collisions, respectively. The $\LcDratio$ production ratios use the corresponding CMS measurements of prompt $\PDz$ production~\cite{HIN16001}.
Centrality bins for $\PbPb$ collisions are given in percentage ranges of the total inelastic hadronic cross section, with the 0--30\% centrality bin corresponding to the 30\% of collisions having the largest overlap of the two nuclei. The values of $\RAA$ are obtained for three centrality intervals: 0--100\%, 0--30\%, and 30--100\%.

\section{The CMS detector}
The central feature of the CMS apparatus is a superconducting solenoid of 6\unit{m} internal diameter, providing a magnetic field of 3.8\unit{T}.
Within the solenoid volume are a silicon tracker, a lead tungstate crystal electromagnetic calorimeter, and a brass and scintillator hadron calorimeter, each composed of a barrel and two endcap sections.  The tracker measures charged particles within the pseudorapidity range $\abseta<2.5$ and the calorimeters record deposited energy for particles with $\abseta<3.0$.
Two forward hadron (HF) calorimeters use steel as an absorber and quartz fibers as the sensitive material. The two HF calorimeters are located 11.2\unit{m} from the interaction region, one on each end, and together they extend the calorimeter coverage from $\abseta=3.0$ to 5.2. Each HF calorimeter consists of 432 readout towers, containing long and short quartz fibers running parallel to the beam, providing information on the shower energy and the relative contribution originating from hadrons versus electrons and photons. A detailed description of the CMS experiment can be found in Ref.~\cite{CMS:detector}.

\section{Event reconstruction and simulated samples}
The total transverse energy deposited in both  HF calorimeters is used to determine the collision centrality in $\PbPb$ collisions and was utilized by the triggers for both data sets included in this analysis~\cite{centrality}.
One trigger selected minimum-bias (MB) events by requiring transverse energy deposits in one (both) HF calorimeters above approximately 1\GeV for $\Pp\Pp$ $(\PbPb)$ collisions. As not all MB events could be saved, an additional trigger selected the more peripheral centrality region of 30--100\% for $\PbPb$ events. The integrated luminosities of $\Pp\Pp$ collisions, $\PbPb$ collisions with centrality 0--100\%, and $\PbPb$ collisions with centrality 30--100\% are 38\nbinv, 44\mubinv, and 102\mubinv, respectively.

The track reconstruction algorithms used in this study for $\Pp\Pp$ and $\PbPb$ collisions are described in Refs.~\cite{ppreconstruct} and~\cite{PbPbreconstruct}, respectively. In $\PbPb$ collisions, minor modifications are made to the $\Pp\Pp$ reconstruction algorithm in order to accommodate the much larger track multiplicities. Tracks are required to have a relative $\pt$ uncertainty of less than 30\% in $\PbPb$ collisions and 10\% in $\Pp\Pp$ collisions. In $\PbPb$ collisions, tracks must also have at least 11 hits and satisfy a stringent fit quality requirement, specifically that the $\chi^2$ per degree of freedom be less than 0.15 times the number of tracker layers with a hit.

For the offline analysis, events must pass selection criteria designed to reject events from background processes (beam-gas interactions and nonhadronic collisions), as described in Ref.~\cite{PbPbreconstruct}. Events are required to have at least one reconstructed primary interaction vertex~\cite{ppreconstruct} with a distance from the center of the nominal interaction region of less than 15\unit{cm} along the beam axis. In addition, in $\PbPb$ collisions, the shapes of the clusters in the pixel detector have to be compatible with those expected from particles produced at the primary vertex location~\cite{ptdistribution}. The $\PbPb$ collision events are also required to have at least three towers in each HF detector with energy deposits of more than 3 GeV per tower. These criteria select $(99 \pm 2)\%$ of inelastic hadronic $\PbPb$ collisions. Fractions above 100\% reflect the possible presence of ultra-peripheral (nonhadronic) collisions in the selected event sample.

Monte Carlo (MC) simulated event samples are used to optimize the selection criteria, calculate the acceptance times efficiency, and estimate the systematic uncertainties.  Proton-proton collisions are generated with \PYTHIA 8.212~\cite{PYTHIA8} tune CUETP8M1~\cite{Eventune}, hereafter referred to as \PYTHIA 8, and includes both prompt and nonprompt $\PcgLp$ baryon events. For the $\PbPb$ MC samples, each \PYTHIA 8 event containing a $\PcgLp$ baryon is embedded into a $\PbPb$ collision event generated with $\HYDJET$ 1.8~\cite{jetquench}, which is tuned to reproduce global event properties such as the charged-hadron $\pt$ spectrum and particle multiplicity. The $\PcgLp \to \Pp \PKm \Pgpp$ decay is performed by $\EVTGEN$ 1.3.0~\cite{Eventgen} through four sub-channels: $\PcgLp \to \Pp \PAK^*(892)^0 \to \Pp \PKm \Pgpp$, $\PcgLp \to \Delta(1232)^{++} \PKm \to \Pp \PKm \Pgpp$, $\PcgLp \to \Lambda(1520) \Pgpp \to \Pp \PKm \Pgpp$, and $\Lcdecay$ (nonresonant), with no modeling of interference between the sub-channels.  All particles are propagated through the CMS detector using the $\GEANTfour$ package~\cite{GEANT4}.

\section{Signal extraction}
\label{sec:signal}
The $\Lcdecay$ candidates are reconstructed by selecting three charged tracks with $\abseta<1.2$ and a net charge of ${+}1$. All tracks must have $\pt >0.7\: (1.0)\,\GeVc$ for $\Pp\Pp$ ($\PbPb$) events. During the invariant mass reconstruction, both possibilities for the mass assignments of the same-sign tracks are considered, while the kaon mass is assigned to the opposite-signed track. Using simulated events, the incorrect assignment was found to produce a broad distribution in the invariant mass (about 30 times the signal width) and is indistinguishable from the combinatorial background.

As the event multiplicities for $\Pp\Pp$ and $\PbPb$ collisions are substantially different, the selection criteria were optimized separately. In the optimization, simulated events in which a reconstructed $\PcgLp$ candidate is matched to a generated $\PcgLp$ baryon are used as the signal sample, and data events from the mass sideband region are used as the background sample.  Requirements are made on three topological and three kinematic variables.  The three topological criteria are: the $\chi^2$ probability of the vertex fit to the three charged tracks making up the $\PcgLp$ candidate, the angle between the $\PcgLp$ candidate momentum and the vector connecting the production and decay vertices in radians ($\alpha$), and the separation between the two vertices.  While more than one collision per bunch crossing is rare in $\PbPb$ collisions, it is common in $\Pp\Pp$ collisions.  Therefore, two-dimensional variables in the transverse plane with respect to the beamline are used for $\alpha$ and decay length in $\Pp\Pp$ collisions, while three-dimensional variables with respect to the primary vertex are used for $\PbPb$ collisions. For the $\PbPb$ events, the topological requirements are $\chi^2$ probability above 20\%, $\alpha<0.1$, and decay length greater than 3.75$\sigma$, where $\sigma$ is the uncertainty in the separation.  For $\Pp\Pp$ events, the corresponding requirements are $\chi^2$ probability above 8\%, $\alpha<0.4$, and decay length greater than 2.25$\sigma$. The kinematic requirements are kaon (proton) $\pt$ divided by the $\PcgLp$ candidate $\pt$ greater than 0.14 (0.28) for all events and pion $\pt$ divided by the $\PcgLp$ candidate $\pt$ greater than 0.12 for $\PbPb$ events.

\begin{figure*}[tb]
\centering
\includegraphics[width=0.32\textwidth]{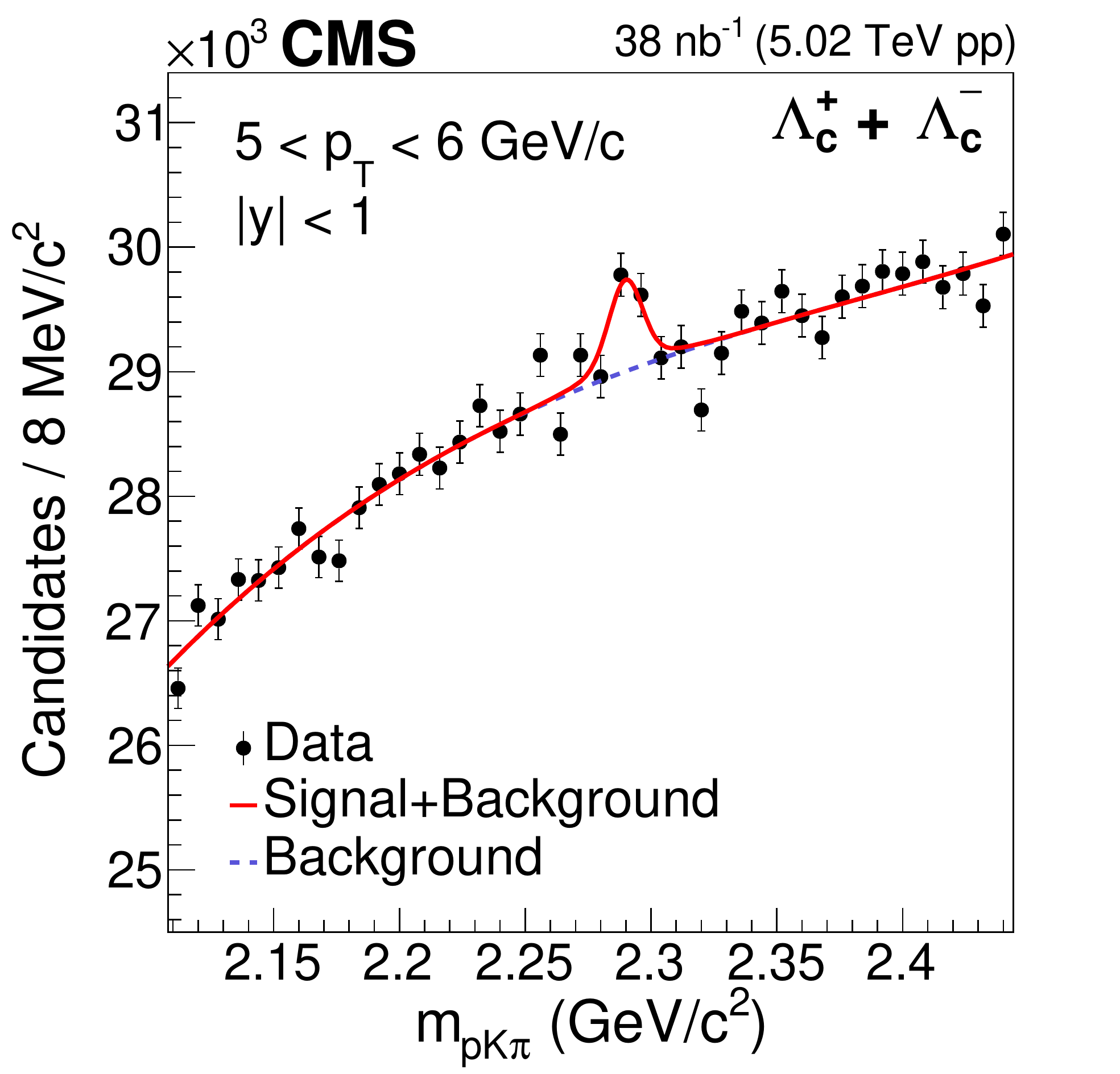}
\includegraphics[width=0.32\textwidth]{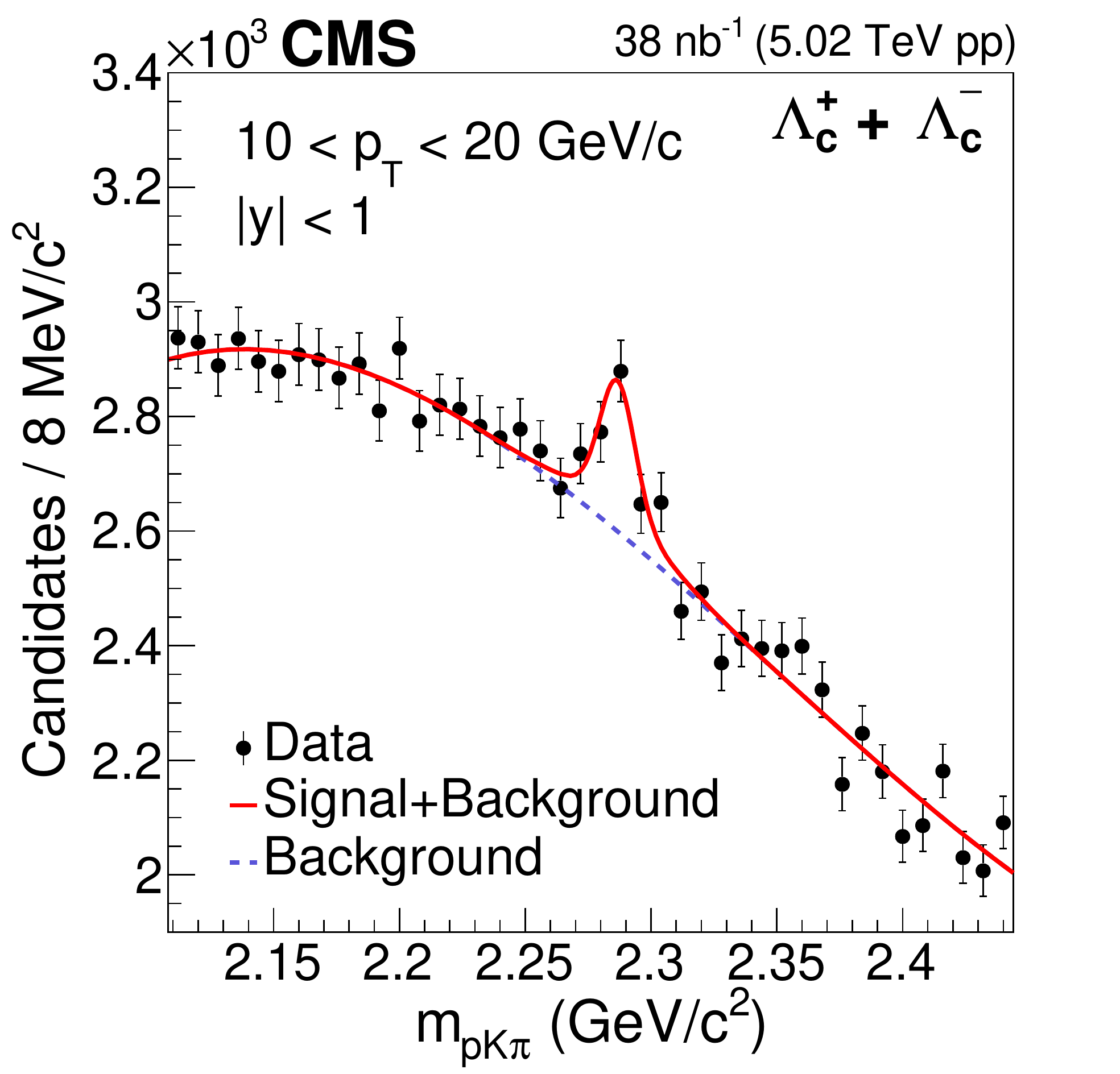}
\includegraphics[width=0.32\textwidth]{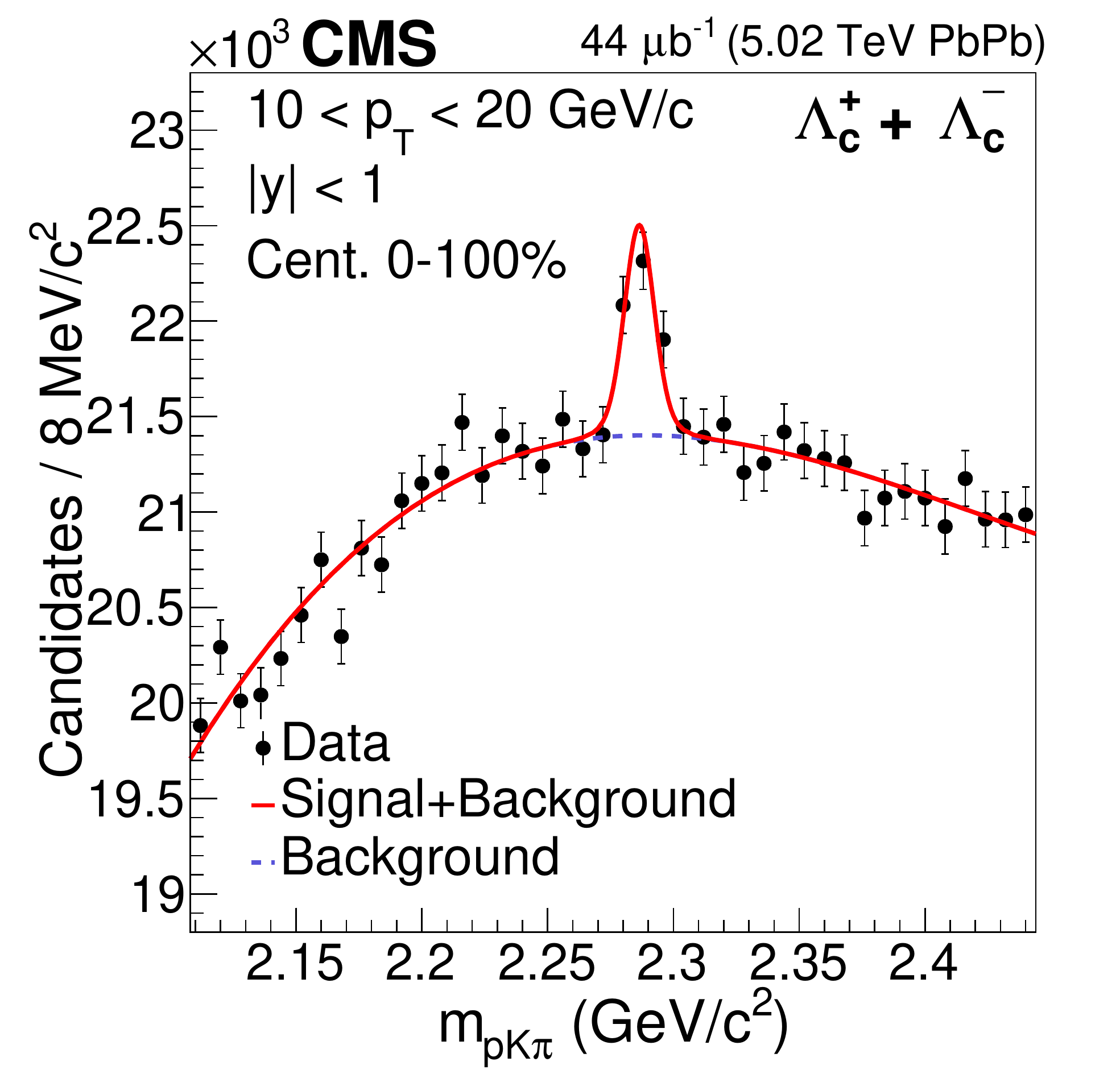}
\caption{Invariant mass distribution of $\PcgLp$ candidates with $\pt$ = 5--6\GeVc (left), 10--20\GeVc(middle) in $\Pp\Pp$ collisions, and $\pt$ = 10--20\GeVc in $\PbPb$ collisions within the centrality range 0--100\% (right). The solid line represents the full fit and the dashed line represents the background component.}
\label{PbPb_default}
\end{figure*}

The $\PcgLp$ baryon yields in each $\pt$ interval are obtained from unbinned maximum likelihood fits to the invariant mass distribution in the range of 2.11--2.45\GeVcc. The signal shape is modeled by the sum of two Gaussian functions with the same mean, but different widths that are fixed on the basis of the simulated signal sample.
One fit parameter scales both widths to accommodate a potential difference in the mass resolution between simulation and data, with the exception of the lowest \pt region (5--6\GeVc) in the $\Pp\Pp$ data, where this parameter was found to cause instability in the fit and the unmodified mass resolution from the simulation was used. The background is modeled with a third-order Chebyshev polynomial. Representative  invariant mass distributions in $\Pp\Pp$ and $\PbPb$ collisions are shown in Fig.~\ref{PbPb_default}.

The $\PcgLp$ baryon differential cross section in $\Pp\Pp$ collisions is defined as:
\begin{linenomath}
\begin{equation}
   \left.\frac{\rd\sigma^{\PcgLp}_{\Pp\Pp}}{\rd\pt}\right|_{\abs{y}<1} = \frac{1}{2\lumi\Delta\pt\mathcal{B}}\left.\frac{N_{\Pp\Pp}^{\PcgLp}|_{\abs{y}<1}}{A\epsilon}\right.,
\end{equation}
\end{linenomath}
where $N_{\Pp\Pp}^{\PcgLp}|_{\abs{y}<1}$ is the $\PcgLp$ yield extracted in each $\pt$ bin, $\lumi$ is the integrated luminosity, $\Delta \pt$ is the width of each $\pt$ bin, $\mathcal{B}$ is the branching fraction of the decay, and $A\epsilon$ is the product of the acceptance and efficiency. The factor of 1/2 accounts for averaging the particle and antiparticle contributions.
The normalized $\PcgLp$ \pt spectrum in $\PbPb$ collisions is defined as:
\ifthenelse{\boolean{cms@external}}{
\begin{multline}
       \frac{1}{\langle T_\text{AA} \rangle}\left.\frac{\rd N^{\PcgLp}_{\PbPb}}{\rd\pt}\right|_{\abs{y}<1} \\= \frac{1}{\langle T_\text{AA} \rangle}\frac{1}{2N_{\text{events}}\Delta\pt\mathcal{B}}\left.\frac{N_{\PbPb}^{\PcgLp}|_{\abs{y}<1}}{A\epsilon}\right.,
\end{multline}
}{
\begin{linenomath}
\begin{equation}
       \frac{1}{\langle T_\text{AA} \rangle}\left.\frac{\rd N^{\PcgLp}_{\PbPb}}{\rd\pt}\right|_{\abs{y}<1} = \frac{1}{\langle T_\text{AA} \rangle}\frac{1}{2N_{\text{events}}\Delta\pt\mathcal{B}}\left.\frac{N_{\PbPb}^{\PcgLp}|_{\abs{y}<1}}{A\epsilon}\right.,
\end{equation}
\end{linenomath}
}
where $N_\text{events}$ is the number of MB events used for the analysis (corrected by the 99\% selection efficiency) and $\langle \taa \rangle$ is the nuclear overlap function, which is equal to the average number of NN binary collisions ($\langle \Ncoll \rangle$) divided by the NN inelastic cross section, and can be interpreted as the NN-equivalent integrated luminosity per heavy ion collision.
The values of $\langle \taa \rangle$, $\langle \Ncoll \rangle$, and the average number of participating nucleons ($\langle \Npart \rangle$) are calculated using a Monte Carlo Glauber model~\cite{glauber}, in which the NN inelastic cross section (70\unit{mb}) is used as an input parameter. The averages of these quantities over the events in the given centrality ranges are listed in Table~\ref{taa}.

\begin{table}[!htbp]
\centering
\topcaption{Summary of the $\langle \Ncoll \rangle$, $\langle \taa \rangle$, and $\langle \Npart \rangle$ values for three $\PbPb$ centrality ranges.}
\renewcommand{\arraystretch}{1.3}
\begin{tabular}{cccc}
\hline
Centrality &  $\langle \taa \rangle [{\mbinv}]$            & $\langle \Npart \rangle$       & $\langle \Ncoll \rangle$  \\ \hline
0--30\%   &  $15.41^{ + 0.33}_{ - 0.47}$ & $270.7^{ + 3.2}_{ - 3.4}$   & $1079^{ + 74}_{ - 78}$\\
30--100\% &  $1.41^{ + 0.09}_{ - 0.06}$  & $46.8^{ + 2.4}_{ - 1.2}$   &$98^{ + 8}_{ - 6}$\\
0--100\%  &  $5.61^{ + 0.16}_{ - 0.19}$ & $114.0^{ + 2.6}_{ - 2.6}$   &$393^{ + 26}_{ - 28}$\\ \hline
\end{tabular}
\label{taa}
\end{table}

The nuclear modification factor $\RAA$ is computed as:
\begin{linenomath}
\begin{equation}
    R_{\mathrm{AA}}(\pt) = \frac{1}{\langle T_\text{AA} \rangle} \left.\frac{\rd N_\text{PbPb}^{\PcgLp}}{\rd\pt}\middle/\frac{\rd\sigma_{\Pp\Pp}^{\PcgLp}}{\rd\pt}\right..
\label{Raa_equation}
\end{equation}
\end{linenomath}
The values of $A\epsilon$ are obtained from MC simulation as a fraction in which the denominator is the number of generated $\PcgLp$ baryons with $\absy<1$ and the numerator is the number of reconstructed $\PcgLp$ candidates that pass the selection criteria and are matched to a generated $\PcgLp$ baryon.  The simulation includes both prompt and nonprompt $\PcgLp$ baryons estimated from \PYTHIA 8 and contains an appropriately weighted combination of decays in the four known sub-channels.  For the $\Pp\Pp$ simulation, the $\pt$ spectrum of the generated $\PcgLp$ baryons is weighted to match a fit to the observed data (iterating until convergence is reached). For $\Pp\Pp$ collisions, $A\epsilon$ increases from 7 to 19\% as $\pt$ increases. As the $\PbPb$ results are given for just one $\pt$ range, an alternative method is used to correct the $\pt$ spectra in simulation.  Under the transverse mass scaling hypothesis ($\mT$ scaling)~\cite{mTscaling}, the $\PcgLp$ baryon $\pt$ spectrum is obtained for the 0--100\% centrality region from the $\PDz$ measurements~\cite{HIN16001} using the function $m^2(\PcgLp) + \pt^2(\PcgLp) = m^2(\PDz) + \pt^2(\PDz)$.  For the $\PbPb$ data set, the centrality distribution in simulation is reweighted to match the data. There is one additional correction applied to $A\epsilon$ for the $\PbPb$ data set.  Previous CMS results have found more suppression for prompt than nonprompt $\PDz$ mesons~\cite{HIN16001,HIN16016}, which can be quantified for $10<\pt<20\GeVc$ as $\Raaratio=1.66\pm0.38$.  As nonprompt baryons tend to have greater \pt and decay farther from the collision point than prompt baryons, the requirement for the decay length significance results in a value of $A\epsilon$ that is larger for nonprompt baryons. Changing the nonprompt fraction to account for the different suppression increases $A\epsilon$ by 15\%. After applying the corrections, $A\epsilon = 5\%$ for $\PbPb$ collisions.

\section{Systematic uncertainties}
Systematic uncertainties arise from the extraction of the raw signal yield, the ability of the MC simulation to reproduce the combined acceptance and efficiency, the branching fraction of the decay mode, and the integrated luminosity. Unless otherwise indicated, systematic uncertainties are combined by adding the individual contributions in quadrature.

The systematic uncertainty in the signal yields is obtained by varying the modeling functions that are used for the signal and background contributions. The background function is changed from the default third- to second- and fourth-order Chebyshev polynomials, with the maximum difference in yield between these two alternative functions and the default fit function taken as the systematic uncertainty. This amounts to 4--10\% and 7--9\% for $\Pp\Pp$ in different $\pt$ bins and $\PbPb$ collisions in three centrality classes, respectively. The default signal model function is the sum of two Gaussian functions with parameters chosen as described in Section~\ref{sec:signal}. For the $\Pp\Pp$ ($\PbPb$) collision data, the alternative model is a triple (single) Gaussian function with similar procedures used for the parameters.  As the signal width is fixed for events in the lowest $\PcgLp$ $\pt$ bin for $\Pp\Pp$ collisions, an additional systematic uncertainty is assessed by varying the width by $\pm$40\%, corresponding to the maximum deviations with respect to the simulation observed in other $\pt$ bins in $\Pp\Pp$ and $\PbPb$ collisions. The uncertainty due to the modeling of the signal is 3--28\% for $\Pp\Pp$ collisions and 2--4\% for $\PbPb$ collisions.

Five sources of systematic uncertainties associated with the MC modeling of the data are evaluated.  The first uncertainty measures the effect of the selection criteria variation.
We define a double ratio as:
\begin{linenomath}
\begin{equation}
    \mathcal{DR} =\left.\frac{N_\text{Data}\text{(varied)}}{N_\text{Data}\text{(nominal)}}\middle/\frac{N_\text{MC}\text{(varied)}}{N_\text{MC}\text{(nominal)}}\right.,
\end{equation}
\end{linenomath}
where $N_\text{Data}\text{(nominal)}$ and $N_\text{Data}\text{(varied)}$ are the yields obtained from data using the default and alternative selection criteria, respectively, and $N_\text{MC}\text{(nominal)}$ and $N_\text{MC}\text{(varied)}$ are the corresponding yields from the simulated events.
For each of the topological selection criteria, the double ratio is evaluated at many different values of the selection criterion.  The specific ranges for $\Pp\Pp$ collision events are ${>}1.5\sigma$ to ${>}6\sigma$, ${>}5\%$ to ${>}45\%$, and ${<}0.1$ to no cut for decay length, vertex fit probability, and $\alpha$, respectively.  The corresponding ranges for $\PbPb$ collision events are  ${>}2.5\sigma$ to ${>}8\sigma$, ${>}5\%$ to ${>}45\%$, and ${<}0.05$ to ${<}0.2$.
For all but the $\alpha$ cut in $\PbPb$ collisions, $\mathcal{DR}$ is plotted as a function of the selection value and fit to a linear function. The systematic uncertainty is taken as the difference between unity and the value of the fitted line at the point where no selection is applied. For the $\alpha$ requirement in $\PbPb$ collisions, the systematic uncertainty is obtained from the biggest differences between unity and the value of $\mathcal{DR}$ from all of the alternative selection values.
Combining the results of the three topological selection criteria systematic uncertainties in quadrature results in uncertainties of 6\% for the $\Pp\Pp$ data set and 19\% for the $\PbPb$ data sets.

The second uncertainty arises from a potential mismodeling of the $\pt$ distribution of $\PcgLp$ baryons because $A\epsilon$ is strongly dependent on the $\PcgLp$ \pt.  In $\Pp\Pp$ collisions, the default $\pt$ shape is derived from the data.  For $\PbPb$ collisions, the default $\pt$ shape is obtained from $\mT$ scaling of the measured $\PDz$ $\pt$ spectrum.  For each data set, two alternative $\pt$ spectra, one from \PYTHIA 8 and one from \PYTHIA 8 with color reconnection (described in Section~\ref{sec:results}) are considered and the maximum deviation in $A\epsilon$ is taken as the systematic uncertainty.  The resulting systematic uncertainty is 0--3\% for $\Pp\Pp$ collisions and 5.2\% for $\PbPb$ collisions.

The third uncertainty arises from imprecise knowledge of the resonant substructure of the $\Pp\PKm\Pgpp$ decay mode~\cite{pdg}.  The calculation of $A\epsilon$ uses the appropriately weighted sum of the four known sub-channels and the systematic uncertainty associated with this is evaluated by determining $A\epsilon$ for each sub-channel and randomly adjusting the weights by the uncertainties of each branching fraction.
The individual values of $A\epsilon$ vary by about ${\pm}30\%$ relative to the average.  The systematic uncertainty is obtained from the standard deviation of a Gaussian fit to the different average $A\epsilon$ values and is 8\% for both $\Pp\Pp$ and $\PbPb$ events.

The fourth uncertainty associated with the MC modeling of the data is the track reconstruction efficiency, which is 4\% for $\Pp\Pp$ collisions~\cite{HIN16001} and 5\% for $\PbPb$ collisions~\cite{HIN16011}.  As there are three tracks in the $\PcgLp$ decay, the corresponding uncertainties on the measured $\pt$ spectra are 12 and 15\% for $\Pp\Pp$ and $\PbPb$, respectively, while for the $\LcDratio$ production ratio, the uncertainties are 4 and 5\%, respectively.

The fifth uncertainty arises from possible mismodeling of the nonprompt component, namely $\PcgLp$ from $\Pb$ hadron decays, in the inclusive $\PcgLp$ sample. The inclusive $A\epsilon$ is the weighted sum of prompt and nonprompt $A\epsilon$ according to the prompt and nonprompt fractions.
As found using the standard \PYTHIA 8 MC sample, the nonprompt $A\epsilon$ is generally 3--4 times larger than the prompt $A\epsilon$ and so an incorrect nonprompt fraction in \PYTHIA 8 will result in an incorrect $A\epsilon$ for the inclusive sample. To evaluate this systematic uncertainty, an alternative method is used to obtain the final result that does not rely on the \PYTHIA 8 prediction for the nonprompt fraction. A generator-only \PYTHIA 8 sample of nonprompt $\PcgLp$ events is reweighted to match the $\pt$-differential $\Pb$ hadron cross section from a fixed-order plus next-to-leading logarithm (FONLL) calculation~\cite{FONLL}. The resulting $\pt$-differential cross section for nonprompt $\PcgLp$ baryons is multiplied by the appropriate luminosity, branching fractions, and $A\epsilon$ for nonprompt $\PcgLp$ events to obtain an estimate of the number of reconstructed nonprompt $\PcgLp$ baryons in each \pt bin.  Subtracting this value from the measured number of reconstructed $\PcgLp$ baryons gives the number of reconstructed prompt $\PcgLp$ baryons. 
These reconstructed prompt yields are then corrected using the prompt $A\epsilon$ as well as luminosity and branching fractions to estimate the \pt-differential cross section for prompt $\PcgLp$ baryons. Finally, the two cross sections give an alternative estimate of the nonprompt fraction in each \pt bin, and therefore an alternative estimate of the weighted inclusive $A\epsilon$ value. The systematic uncertainty is taken as the difference between the nominal and alternative $A\epsilon$ values. The nonprompt fraction for events passing the $\Pp\Pp$ selection criteria is found to be 28--34\% for the nominal scenario (\PYTHIA 8 only) and 4--7\% for the alternative method, with higher values associated with larger $\PcgLp$ \pt. The resulting systematic uncertainty varies by only $\pm 1\%$ as a function of \pt so an average value of 18\% is used for all \pt bins. The same method is applied to the $\PbPb$ data set, where the systematic uncertainty is found to be 25\% as a result of the more stringent selection criteria.  For $\PbPb$ collisions, an additional systematic uncertainty is assessed by taking the difference between applying and not applying the correction for different values of $\RAA$ for nonprompt and prompt $\PcgLp$ baryons as discussed in Section~\ref{sec:signal}, raising the systematic uncertainty to 29\%.

The overall $\Lcdecay$ branching fraction uncertainty is 5.3\%~\cite{pdg}.  The uncertainties in the integrated luminosity in $\Pp\Pp$ collisions and the MB selection efficiency in $\PbPb$ collisions are 2.3\%~\cite{lumi_pp} and 2.0\%~\cite{PbPbreconstruct}, respectively. The uncertainties in $\taa$ are listed in Table~\ref{taa}.

For the measurement of the $\pt$ spectra, the uncertainties associated with the $\Lcdecay$ branching fraction and subresonant contributions, the luminosity and MB selection efficiency, and the nonprompt fraction contribute only to the overall normalization and are labeled global uncertainties.  Adding these contributions in quadrature yields global uncertainties of 21\% (31\%) for $\Pp\Pp$ $(\PbPb)$ collisions.  In measuring the nuclear modification factor $\RAA$, the uncertainties associated with the branching fraction and subresonant contributions cancel and the nonprompt fraction uncertainty partially cancels.  In calculating the $\LcDratio$ production ratio, the uncertainties associated with $\PDz$ from the yield extraction, selection criteria efficiency, and $\pt$ shape are obtained from Ref.~\cite{HIN16001}, while the uncertainties in the integrated luminosity in $\Pp\Pp$ collisions and the MB selection efficiency in $\PbPb$ collisions cancel.

\section{Results and discussion}
\label{sec:results}

\begin{figure}[htb]\centering
 \includegraphics[width=\cmsFigWidth]{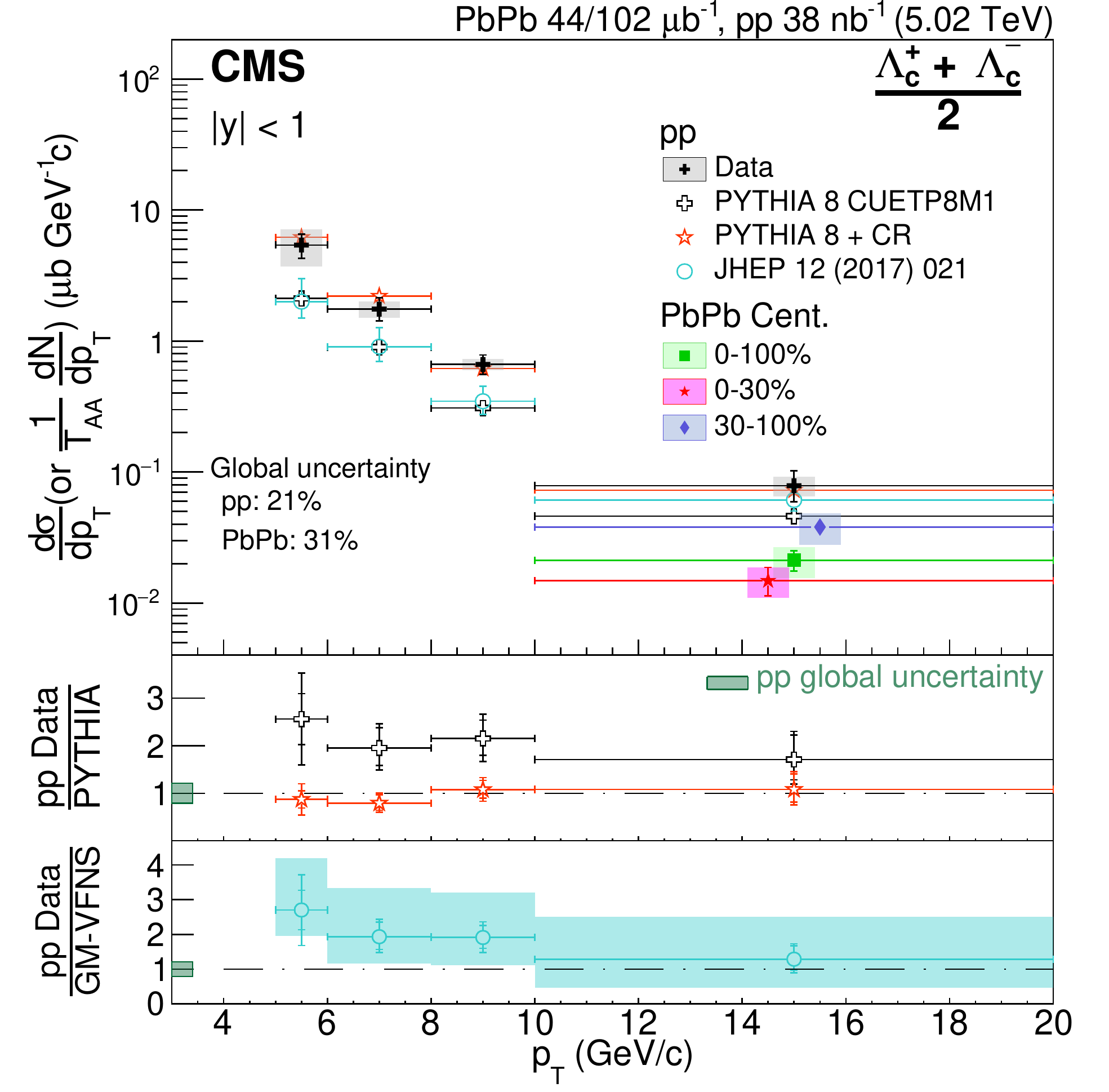}
 \caption{The \pt-differential cross sections for inclusive $\PcgLp$ production in $\Pp\Pp$ collisions and the $T_\text{AA}$-scaled yields for three centrality regions of $\PbPb$ collisions. The boxes and error bars represent the systematic and statistical uncertainties, respectively. The $\PbPb$ data points are shifted in the horizontal axis for clarity. Predictions for $\Pp\Pp$ collisions are displayed for \PYTHIA 8 with the CUETP8M1 tune (open crosses), \PYTHIA 8 with color reconnection~\cite{CRmodes} (open stars), and GM-VFNS~\cite{gmvfns_model} (open circles labeled ``JHEP 12 (2017) 021'') along with ratios to the data in the lower two panels. The \PYTHIA 8 (GM-VFNS) predictions are for inclusive (prompt) $\PcgLp$ production. The error bars on the GM-VFNS prediction account for the scale variation uncertainty. The lower panels show the data-to-prediction ratio for $\Pp\Pp$ collisions with inner and outer error bars corresponding to the statistical and total uncertainty in the data, respectively, and the shaded box at unity indicating the 21\% normalization uncertainty. The shaded boxes in the bottom panel represent the GM-VFNS uncertainty.
}\label{yield}\end{figure}

Figure~\ref{yield} shows the \pt-differential cross section of inclusive $\PcgLp$ baryon production in $\Pp\Pp$ collisions for the range of $5<\pt<20\GeVc$ and the \taa-scaled yields in $\PbPb$ collisions for the range of $10<\pt<20\GeVc$, for three centrality classes. The 21\% (31\%) normalization uncertainty for the $\Pp\Pp$ ($\PbPb$) results is not included in the boxes representing the systematic uncertainties for each data point.
While the shape of the $\pt$ distribution in $\Pp\Pp$ collisions is consistent with the inclusive production calculation from \PYTHIA 8 using tune CUETP8M1 and activating the ``SoftQCD:nondiffractive'' processes, the data are systematically higher.
The hadronization in \PYTHIA 8 can be modified by adding a color reconnection (CR) mechanism in which the final partons in the string fragmentation are considered to be color connected in such a way that the total string length becomes as short as possible~\cite{CRmodes}. The calculations using the recommended color reconnection model from Ref.~\cite{CRmodes} are consistent with our \pt-differential cross section in $\Pp\Pp$ collisions.
The \pt-differential cross section in $\Pp\Pp$ collisions is also compared to the GM-VFNS perturbative QCD calculations~\cite{gmvfns_model}, which includes only prompt $\PcgLp$ baryon production.  The GM-VFNS prediction is significantly below our data for $\pt<10\GeVc$, similar to the difference found by ALICE~\cite{ALICE_pp}.  \PYTHIA 8 predicts that 8--15\% of generated $\PcgLp$ baryons arise from $\Pb$ hadrons, with the low (high) value corresponding to the $\PcgLp$ \pt interval $5<\pt<6\GeVc$ $(10<\pt<20\GeVc)$.
Therefore, accounting for the effects of nonprompt $\PcgLp$ production will only marginally reduce the disagreement with the GM-VFNS prediction.

\begin{figure}[!htb]\centering
 \includegraphics[width=\cmsFigWidth]{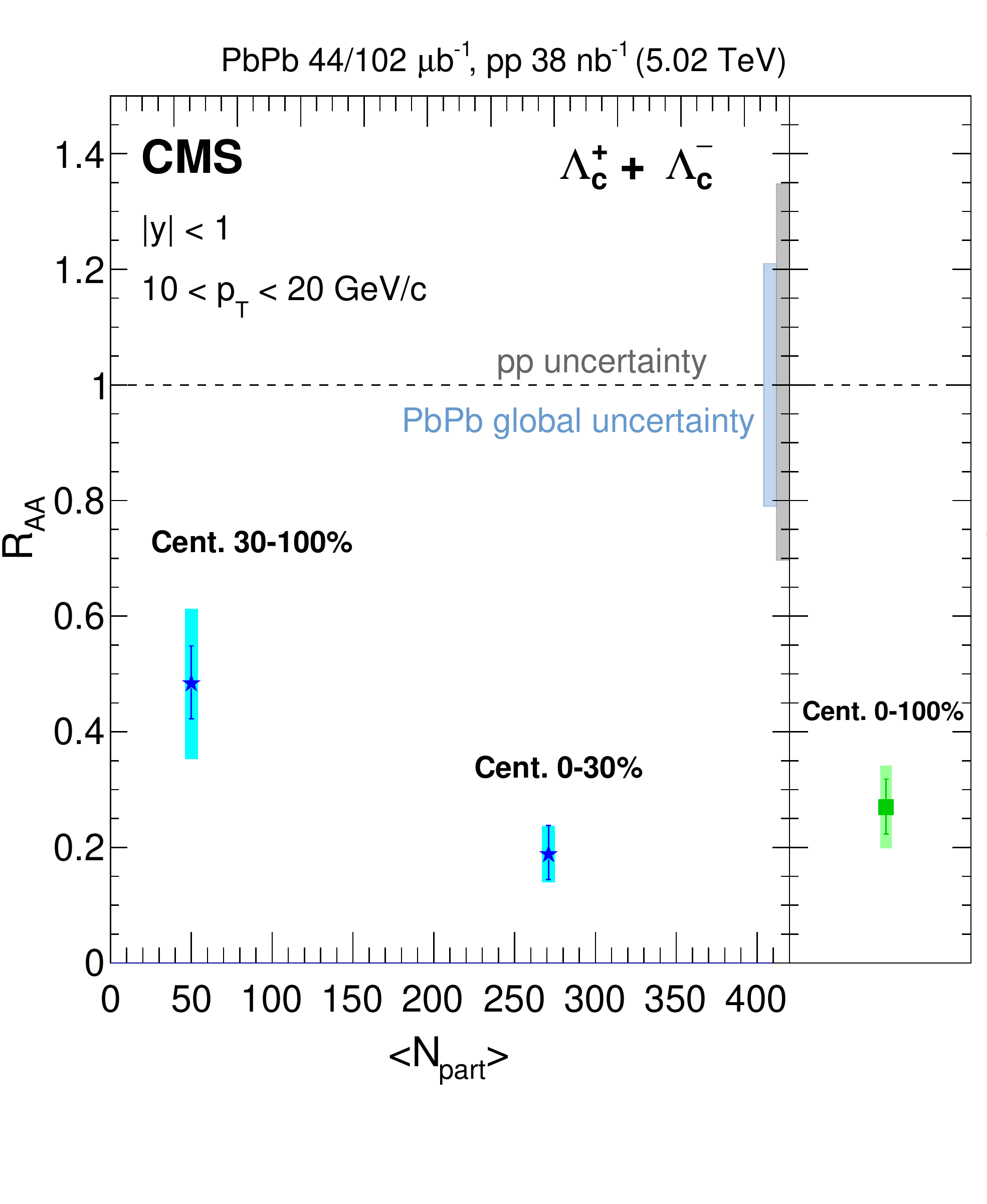}
 \caption{The nuclear modification factor $\RAA$ versus $\langle \Npart \rangle$ for inclusive $\PcgLp$ production. The error bars represent the $\PbPb$ yield statistical uncertainties.  The boxes at each point include the $\PbPb$ systematic uncertainties associated with the signal extraction, $\pt$ spectrum, selection criteria, track reconstruction, and \taa.  The band at unity labeled $\Pp\Pp$ uncertainty includes these same uncertainties for the $\Pp\Pp$ data (except for \taa) plus the uncertainties in $\Pp\Pp$ yield and luminosity.  The band at unity labeled $\PbPb$ includes the uncertainty from the nonprompt fraction (accounting for a partial cancelation between $\Pp\Pp$ and $\PbPb$) and MB selection efficiency.}
 \label{Raa_plot}\end{figure}

The nuclear modification factor $\RAA$ for inclusive $\PcgLp$ baryons in the $\pt$ range 10--20\GeVc is shown in Fig.~\ref{Raa_plot} as a function of the number of participating nucleons $\langle \Npart \rangle$ for $\PbPb$ collisions.  The results suggest that $\PcgLp$ is suppressed in $\PbPb$ collisions for $\pt>10\GeVc$, but no conclusion can be drawn because of the large uncertainties. The difference in $\RAA$ values between the 0--30\% and 30--100\% centrality ranges is consistent with an enhanced suppression in the more central $\PbPb$ collisions.

\begin{figure}[htb]\centering
 \includegraphics[width=\cmsFigWidthHalf]{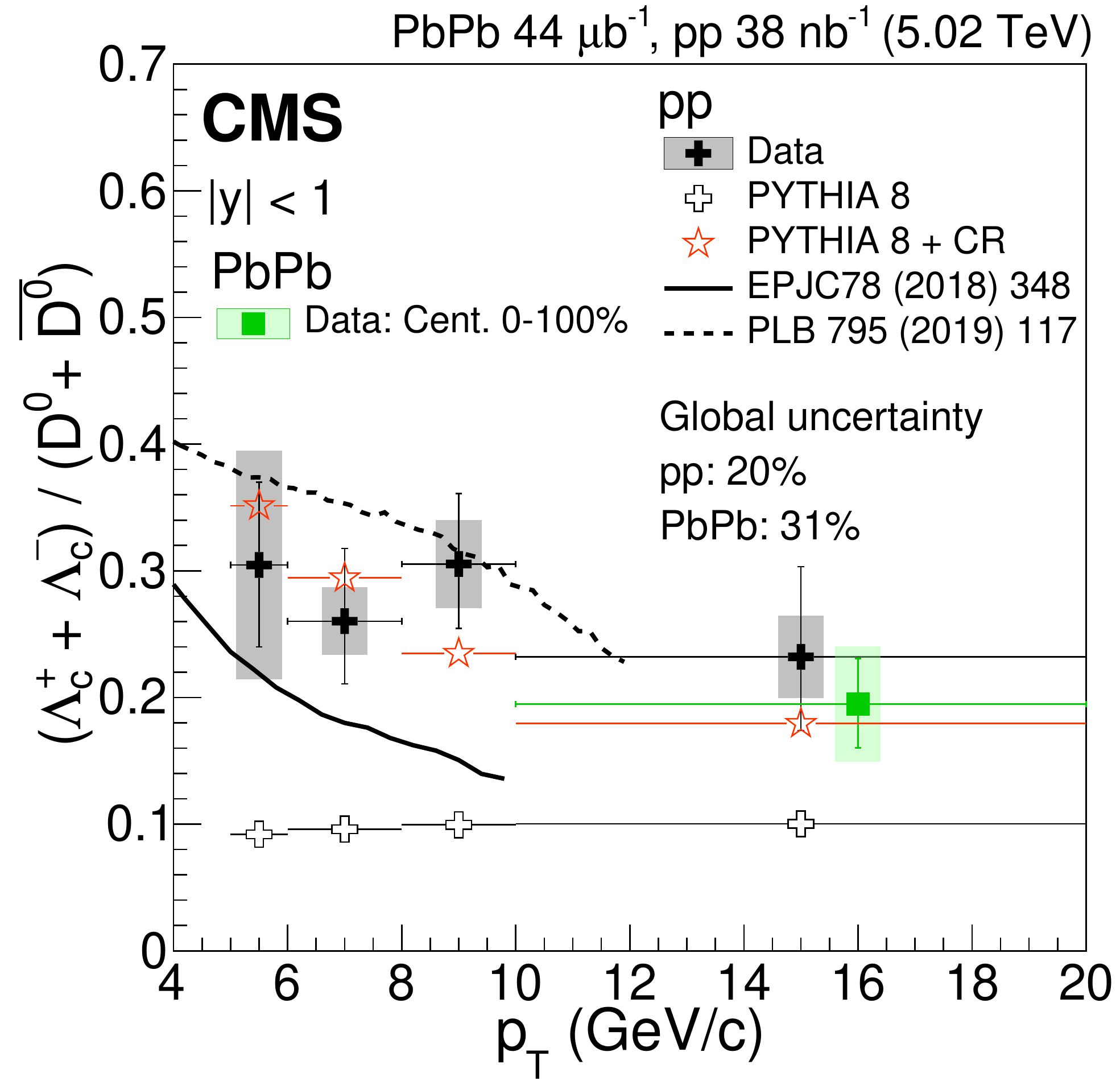}
 \caption{The ratio of the production cross sections of inclusive $\PcgLp$ to prompt $\PDz$ versus \pt from $\Pp\Pp$ collisions as well as 0--100\% centrality $\PbPb$ collisions. The boxes and error bars represent the systematic and statistical uncertainties, respectively. The  $\PbPb$ data point is shifted in the horizontal axis for clarity. The 20 and 31\% normalization uncertainties in $\Pp\Pp$ and $\PbPb$ collisions, respectively, are not included in the boxes representing the systematic uncertainties for each data point.
The open crosses and open stars represent the predictions of \PYTHIA 8 with the CUETP8M1 tune and with color reconnection~\cite{CRmodes}, respectively. The solid and dashed lines are the calculations for prompt $\PcgLp$ over prompt $\PDz$ production ratio from Ref.~\cite{Greco_publication} and Ref.~\cite{Rapp_ppresult}, respectively. All predictions are for $\Pp\Pp$ collisions.}
 \label{LcD0_plot}\end{figure}

Figure~\ref{LcD0_plot} shows the $\LcDratio$ production ratio as a function of $\pt$ for $\Pp\Pp$ collisions and $\PbPb$ collisions in the centrality range 0--100\%.
The production ratio found from $\Pp\Pp$ collisions is similar in shape versus $\pt$ but about three times larger in magnitude compared to the calculation from \PYTHIA 8.212 tune CUETP8M1.
Results using the Monash 2013~\cite{Monashtune} tune are found to be consistent with those from the CUETP8M1 tune.
Besides providing a reasonable description of $\PcgLp$ baryon \pt-differential cross sections, Fig.~\ref{LcD0_plot} shows that calculations using a color reconnection model are consistent with our results for the $\LcDratio$ production ratio in $\Pp\Pp$ collisions.

The $\Pp\Pp$ data are also compared with two predictions which are for the prompt $\PcgLp$ over $\PDz$ production ratio.
Calculations using a model that includes both coalescence and fragmentation in $\Pp\Pp$ collisions~\cite{Greco_publication} are shown in Fig.~\ref{LcD0_plot} by the solid line.
Compared to the data, this model predicts a stronger dependence on $\pt$ and underestimates the measurements.
Another recent model~\cite{Rapp_ppresult} attempts to use a statistical hadronization approach to explain the large $\LcDratio$ production ratio as arising from $\PcgLp$ baryons that are produced from the decay of excited charm baryon states not included in Ref.~\cite{pdg} and are therefore not included in the hadronization simulation in \PYTHIA 8. The prediction of this model, also shown in Fig.~\ref{LcD0_plot} by the dashed line, provides a reasonable description of the data for $\pt<10\GeVc$.

While the ALICE results indicate an enhancement in the $\LcDratio$ production ratio in the $\pt$ range of 6--12\GeVc for $\PbPb$~\cite{ALICE_PbPb} compared to $\Pp$Pb and $\Pp\Pp$ collisions, the CMS $\PbPb$ measurement in the $\pt$ range 10--20\GeVc is consistent with the $\Pp\Pp$ result.
This lack of an enhancement may suggest that there is no significant contribution from the coalescence process for $\pt>10\GeVc$ in $\PbPb$ collisions.

\section{Summary}

The \pt-differential cross sections of $\PcgLp$ baryons, including both prompt and nonprompt contributions, have been measured in $\Pp\Pp$ and $\PbPb$ collisions at a nucleon-nucleon center-of-mass energy of 5.02\TeV. The shape of the $\pt$ distribution in $\Pp\Pp$ collisions is well described by the \PYTHIA 8 event generator. A hint of suppression of $\PcgLp$ production for $10<\pt<20\GeVc$ is observed in $\PbPb$ when compared to $\Pp\Pp$ data, with central $\PbPb$ events showing stronger suppression.
This is consistent with the suppression observed in $\PDz$ meson measurements, which is understood to originate from the strong interaction between the charm quark and the quark-gluon plasma.
The $\LcDratio$  production ratios in $\Pp\Pp$ collisions are consistent with a model obtained by adding color reconnection in hadronization to \PYTHIA 8, and also with a model that includes enhanced contributions from the decay of excited charm baryons.
The $\LcDratio$ production ratios in $\Pp\Pp$ and $\PbPb$ collisions for $\pt$ = 10--20\GeVc are found to be consistent with each other. These two observations may suggest that the coalescence process does not play a significant role in $\PcgLp$ baryon production in this $\pt$ range.

\begin{acknowledgments}
We thank V. Greco for providing the theoretical calculations of the $\LcDratio$ production ratios used for comparisons with our measurements in $\Pp\Pp$ collisions.

We congratulate our colleagues in the CERN accelerator departments for the excellent performance of the LHC and thank the technical and administrative staffs at CERN and at other CMS institutes for their contributions to the success of the CMS effort. In addition, we gratefully acknowledge the computing centres and personnel of the Worldwide LHC Computing Grid for delivering so effectively the computing infrastructure essential to our analyses. Finally, we acknowledge the enduring support for the construction and operation of the LHC and the CMS detector provided by the following funding agencies: BMBWF and FWF (Austria); FNRS and FWO (Belgium); CNPq, CAPES, FAPERJ, FAPERGS, and FAPESP (Brazil); MES (Bulgaria); CERN; CAS, MoST, and NSFC (China); COLCIENCIAS (Colombia); MSES and CSF (Croatia); RPF (Cyprus); SENESCYT (Ecuador); MoER, ERC IUT, PUT and ERDF (Estonia); Academy of Finland, MEC, and HIP (Finland); CEA and CNRS/IN2P3 (France); BMBF, DFG, and HGF (Germany); GSRT (Greece); NKFIA (Hungary); DAE and DST (India); IPM (Iran); SFI (Ireland); INFN (Italy); MSIP and NRF (Republic of Korea); MES (Latvia); LAS (Lithuania); MOE and UM (Malaysia); BUAP, CINVESTAV, CONACYT, LNS, SEP, and UASLP-FAI (Mexico); MOS (Montenegro); MBIE (New Zealand); PAEC (Pakistan); MSHE and NSC (Poland); FCT (Portugal); JINR (Dubna); MON, RosAtom, RAS, RFBR, and NRC KI (Russia); MESTD (Serbia); SEIDI, CPAN, PCTI, and FEDER (Spain); MOSTR (Sri Lanka); Swiss Funding Agencies (Switzerland); MST (Taipei); ThEPCenter, IPST, STAR, and NSTDA (Thailand); TUBITAK and TAEK (Turkey); NASU and SFFR (Ukraine); STFC (United Kingdom); DOE and NSF (USA).

\hyphenation{Rachada-pisek} Individuals have received support from the Marie-Curie programme and the European Research Council and Horizon 2020 Grant, contract Nos.\ 675440, 752730, and 765710 (European Union); the Leventis Foundation; the A.P.\ Sloan Foundation; the Alexander von Humboldt Foundation; the Belgian Federal Science Policy Office; the Fonds pour la Formation \`a la Recherche dans l'Industrie et dans l'Agriculture (FRIA-Belgium); the Agentschap voor Innovatie door Wetenschap en Technologie (IWT-Belgium); the F.R.S.-FNRS and FWO (Belgium) under the ``Excellence of Science -- EOS" -- be.h project n.\ 30820817; the Beijing Municipal Science \& Technology Commission, No. Z181100004218003; the Ministry of Education, Youth and Sports (MEYS) of the Czech Republic; the Lend\"ulet (``Momentum") Programme and the J\'anos Bolyai Research Scholarship of the Hungarian Academy of Sciences, the New National Excellence Program \'UNKP, the NKFIA research grants 123842, 123959, 124845, 124850, 125105, 128713, 128786, and 129058 (Hungary); the Council of Science and Industrial Research, India; the HOMING PLUS programme of the Foundation for Polish Science, cofinanced from European Union, Regional Development Fund, the Mobility Plus programme of the Ministry of Science and Higher Education, the National Science Center (Poland), contracts Harmonia 2014/14/M/ST2/00428, Opus 2014/13/B/ST2/02543, 2014/15/B/ST2/03998, and 2015/19/B/ST2/02861, Sonata-bis 2012/07/E/ST2/01406; the National Priorities Research Program by Qatar National Research Fund; the Ministry of Science and Education, grant no. 3.2989.2017 (Russia); the Programa Estatal de Fomento de la Investigaci{\'o}n Cient{\'i}fica y T{\'e}cnica de Excelencia Mar\'{\i}a de Maeztu, grant MDM-2015-0509 and the Programa Severo Ochoa del Principado de Asturias; the Thalis and Aristeia programmes cofinanced by EU-ESF and the Greek NSRF; the Rachadapisek Sompot Fund for Postdoctoral Fellowship, Chulalongkorn University and the Chulalongkorn Academic into Its 2nd Century Project Advancement Project (Thailand); the Welch Foundation, contract C-1845; and the Weston Havens Foundation (USA).

\end{acknowledgments}
\bibliography{auto_generated}

\cleardoublepage \appendix\section{The CMS Collaboration \label{app:collab}}\begin{sloppypar}\hyphenpenalty=5000\widowpenalty=500\clubpenalty=5000\input{HIN-18-009-authorlist.tex}\end{sloppypar}
\end{document}

%% file: HIN-18-009-authorlist.tex
\vskip\cmsinstskip
\textbf{Yerevan Physics Institute, Yerevan, Armenia}\\*[0pt]
A.M.~Sirunyan$^{\textrm{\dag}}$, A.~Tumasyan
\vskip\cmsinstskip
\textbf{Institut f\"{u}r Hochenergiephysik, Wien, Austria}\\*[0pt]
W.~Adam, F.~Ambrogi, T.~Bergauer, J.~Brandstetter, M.~Dragicevic, J.~Er\"{o}, A.~Escalante~Del~Valle, M.~Flechl, R.~Fr\"{u}hwirth\cmsAuthorMark{1}, M.~Jeitler\cmsAuthorMark{1}, N.~Krammer, I.~Kr\"{a}tschmer, D.~Liko, T.~Madlener, I.~Mikulec, N.~Rad, J.~Schieck\cmsAuthorMark{1}, R.~Sch\"{o}fbeck, M.~Spanring, D.~Spitzbart, W.~Waltenberger, J.~Wittmann, C.-E.~Wulz\cmsAuthorMark{1}, M.~Zarucki
% ...
\vskip\cmsinstskip
\dag: Deceased\\
1:  Also at Vienna University of Technology, Vienna, Austria\\
% ...
